\documentclass[10.5pt, aps,prl,twocolumn,nofootinbib]{revtex4-1}

\usepackage{graphicx}

\begin{document}

\title{Can Higgs Inflation be Saved with High-scale Supersymmetry ?}

\author{Sibo Zheng}
\affiliation{Department of Physics, Chongqing University, Chongqing 401331, P. R. China}

\begin{abstract}
It is shown whether Higgs inflation can be saved with high-scale supersymmetry critically 
depends on the magnitude of non-minimal coupling constant $\xi$.
For small $\xi \leq 500$, 
the threshold correction at scale $M_{P}/\xi$ is constrained in high precision.
Its magnitude is in the narrow range of $(-0.03, -0.02)$  and $(-0.05, -0.04)$ for the wino and higgsino/singlino dark matter, respectively.
While in the large $\xi$-region with $\xi \geq 10^{4}$, 
such high-scale supersymmetry is excluded by too large threshold correction as required by Higgs inflation.
\end{abstract}
\maketitle

\section{I.~Introduction }
The collider facilities such as the Large Hadron Collider (LHC) and astrophysical experiments 
are two main tools for exploring new physics beyond the standard model (SM).
Recently, the Plank Collaboration,
designed to detect the cosmic microwave background (CMB) temperature anisotropy and polarization,
has reported the latest value of tensor-to-scalar ratio $r\leq 0.11$ at $95\%$ CL \cite{1502.01589, 1502.02114}.
This data excludes a few well known inflation models such as quadratic inflation,
but still allows some simple examples such as Starobinsky-like inflation \cite{Starobinsky}, 
$\alpha$-attractor inflation \cite{1502.07733} and Higgs inflation \cite{0710.3755}.

Among these survivors, Higgs inflation is rather special due to two considerations.
Firstly, there is only one new parameter $\xi$ in this model in the Lagrangian in this model,
which reads as in the Jordan frame,
\begin{eqnarray}\label{Lagrangian}
\mathcal{L}_{J}=\frac{M^{2}+\xi h^{2}}{2} R-\frac{1}{2} \left(\partial h\right)^{2}-\frac{\lambda_{H}}{4}\left(h^{2}-\upsilon^{2}\right)^{2}.
\end{eqnarray}
Here, $\xi$  measures the non-minimal coupling  between Higgs scalar $h$ and gravity.
It was found that the Higgs scalar potential $V_{E}(h)$ in the Einstein frame rapidly approaches to a plateau potential in the large field region $h> M_{P}/\sqrt{\xi}$, 
and the model predicts that for e-foldings number $N=60$, 
\begin{eqnarray}\label{nr}
n_{s}\simeq 0.970, ~~~~r\simeq 0.0033,
\end{eqnarray}
which is in perfect agreement with the Plank 2015 data \cite{1502.01589,1502.02114}.
In terms of the present cosmological data 
$M$ in Eq.(\ref{Lagrangian}) approximates to the reduced Plank mass $M_{P}=2.4\times 10^{18}$ GeV 
for our present universe.
Moreover, unlike Starobinsky-like inflation or $\alpha$-attractor inflation,
it is obvious that Higgs inflation is of special interest from the viewpoint of particle physics.

Unfortunately,  there are two subjects against the Higgs inflation.
The first subject shows that the SM is not valid above scale $M_{P}/\xi$ \cite{0902.4465,0903.0355,1002.2730}.
It implies that the plateau potential $V(h)$ at large field value $h> M_{P}/\sqrt{\xi}$ may be significantly modified by the ultra-violet completion
so that realistic inflation driven by the Higgs field can not occur at all.
It is proposed in \cite{1008.5157} that a scale symmetry in the ultra-violet completion may be the prescription to this problem.
The second subject arises since the discovery of the Higgs mass $m_{h}=125.5\pm 0.5$ GeV \cite{Higgsmass1, Higgsmass2, Higgsmass3},
which implies that the SM Higgs quartic coupling $\lambda_{H}$ at the weak scale is not large enough 
so that $\lambda_{H}<0$ above the high-energy scale $10^{9}-10^{11}$ GeV.
The uncertainty about this critical scale mainly arises from the the uncertainty of top quark mass,
as the renormalization group equation (RGE) for $\lambda_{H}$ is very sensitive to the top Yukawa coupling.
Obviously,  positive $\lambda_{H}$ is required during inflationary epoch. 
This problem can be solved through introducing some new fields into the SM, 
which give rise to either correct threshold correction to $\lambda_{H}$ or slowing down the RG evaluation for $\lambda_{H}$ along high energy scale, see, .e.g., \cite{1203.0237,1405.7331}.

In this letter, we consider saving the Higgs inflation with high-scale supersymmetry (SUSY).
High-scale SUSY can provide the observed Higgs mass at the LHC and stable DM candidate \cite{1409.2939}.
However,  the detection of such models at colliders is unpromising.
In this sense the astrophysical probe is an important direction.
Fortunately,  they can be probed directly or indirectly 
in the light of the cosmological observations on the early universe \cite{1109.0292,1409.7462}.

We identify the cut-off scale $M_{P}/\xi $ 
as the typical mass scale $\tilde{m}$ in the SUSY mass spectrum. 
Above the scale $\tilde{m}=M_{P}/\xi $
supergravity is the natural ultra-violet (UV) completion, 
and for the matter content same as the next-to-minimal SUSY model (NMSSM) 
it may maintain the plateau potential  \cite{1008.2942}. 
For earlier discussions, see \cite{1009.2276, 0912.2718, 1004.0712}.
On the other hand, the SUSY dark matter (DM), whose mass is around the weak scale, 
is the natural choice on the new fields added to the SM below the cut-off scale $\tilde{m}$.
In this paper, we consider two possibilities \cite{1409.2939},
where wino $\tilde{w}$ and higgsino/singlino mixing state serves as the DM, respectively.

The paper is organized as follows.
In the next section, we discuss the RGE for $\lambda_{H}$ in two classes of high-scale SUSY below scale $\tilde{m}$.
In Section III, we discuss embedding the Higgs inflation to supergravity,
where we uncover the constraint on parameters $\lambda_{H}$ and $\xi$ at the end of inflation $h_{\text{end}}=M_{P}/\sqrt{\xi}$ arising from the present cosmological data.
In section IV, we discuss the constraint on the SUSY mass spectrum
in terms of the constraint on the threshold correction to $\lambda_{H}$.
Finally, we discuss our results in section V.

\section{II.~RGE for $\lambda_{H}$ Below Scale $M_{P}/\xi$}

Below the scale $\tilde{m}$,
the effective theory is described by the SM together 
with either wino-like or higgsino/singlino-like DM.
In this section we use the 2-loop RGEs for relevant couplings for our analysis.
We refer the reader to appendix in our previous work \cite{1409.2939},
where the 2-loop RGEs for SM gauge and Yukawa couplings, and Higgs quartic coupling
in models of SM$+\tilde{w}$  and SM$+\tilde{h}/\tilde{s}$ are explicitly shown.
We ignore other SM Yukawa couplings in the following discussion.
For the calculation of beta functions, see the references therein.

\begin{figure}
\includegraphics[width=0.45\textwidth]{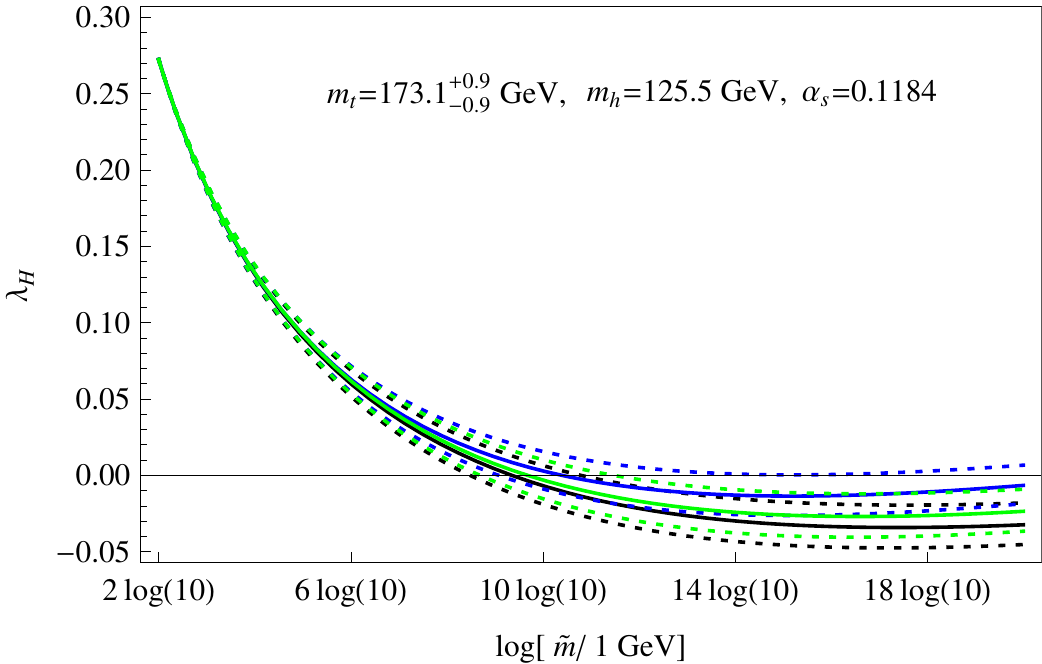}
\caption{Two-loop RG running for $\lambda_{H}$ in the SM (black), SM$+\tilde{w}$ (blue) and SM$+\tilde{h}/\tilde{s}$ (green), respectively. 
The solid lines correspond to the central value $m_{t}=173.1$ GeV, $m_{h}=125.5$ GeV and $\alpha_{s}(m_{Z})=0.1184$. The dotted lines show the uncertainty due to the top quark mass with 1$\sigma$ deviation. In the model of SM$+\tilde{h}/\tilde{s}$, we have chosen small value $g_{\lambda}(m_{Z})=0.2$. See the text for the reason for this choice.}
\label{lambda1}
\end{figure}

Fig.\ref{lambda1} shows the RG running for $\lambda_{H}$ in high-scale SUSY  discussed in this letter, 
which is modified by the SUSY DM (wino or higgisno-singlino mixing state) with mass of weak scale in compared with the SM.
In this figure, black, blue and green curves corresponds to the SM, SM$+\tilde{w}$, and SM$+\tilde{h}/\tilde{s}$, respectively. 
The solid lines refer to the central value $m_{t}=173.1$ GeV, $m_{h}=125.5$ GeV and $\alpha_{s}(m_{Z})=0.1184$,
while the dotted lines show the uncertainty due to the top quark mass with 1$\sigma$ deviation. 
From Fig.\ref{lambda1} we observe that more positive $\lambda_{H}$ is obtained in high-scale SUSY with
wino DM for smaller top quark pole mass $m_{t}=172.2$ GeV along RG running to scale $M_{P}/\xi$. 
This is mainly due to the correction to the SM beta function $\beta_{\lambda_{H}}$ induced by the wino DM.

Unlike the case with wino DM, 
the sign of the correction induced by higgsino/singlino DM is only positive 
when model parameter $g_{\lambda}$ is small.
Otherwise, for large $g_{\lambda}\geq 0.4$ the sign of the correction will reverse 
and $\lambda_{H}$ will be negative more rapidly along RG running to scale $M_{P}/\xi$.
This implies that the correction in this case is upper bounded.

\begin{figure}
\includegraphics[width=0.45\textwidth]{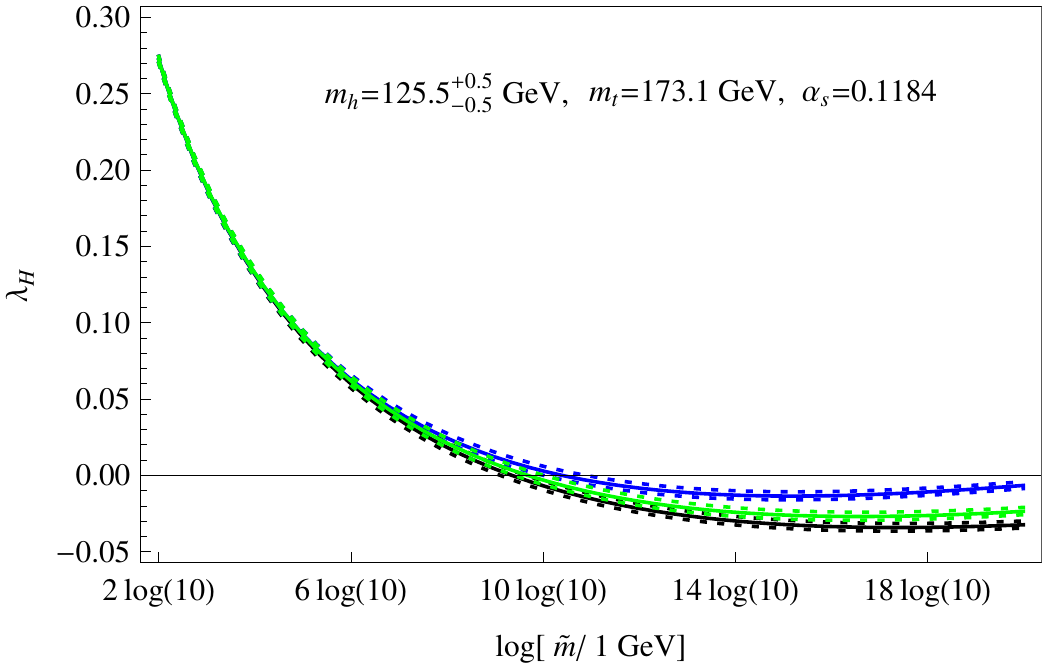}
\caption{Same as Fig.\ref{lambda1} but with $m_{t}=173.1$ GeV and $m_{h}=125.5\pm 0.5$ GeV.}
\label{lambda2}
\end{figure}

Fig.\ref{lambda2} shows the sensitivity of the RG running for $\lambda_{H}$ to the Higgs mass. 
One finds that for about $\sim 0.5$ GeV deviation to the central value $m_{h}=125.5$ GeV 
the correction in each case is tiny.
Among the three models, $\lambda_{H}$ approaches to zero mostly in the case with wino DM, 
similarly to what Fig.\ref{lambda1} indicates.

In summary,  for about $1\sigma$ deviation to the central value of top quark mass, 
$\lambda_{H}$ is tuned to be more positive in high-scale SUSY 
with the low-energy theory below scale $M_{P}/\xi$ 
described by either SM$+\tilde{w}$ or SM$+$higgsino/singlino DM in compared with the SM.
As we will see in the next section,
more positive $\lambda_{H}$ at scale $M_{P}/\xi$ is favored by the Higgs inflation.
Because the threshold correction at scale $M_{P}/\xi$,
may be unable to tune $\lambda_{H}$ into a  positive parameter above this scale as required by Higgs inflation.

\section{III.~Embedding Higgs Inflation into Supergravity}
We proceed to discuss the constraint arising from the condition of plateau potential for Higgs scalar 
in the context of supergravity, which is automatically the ultra-violet completion.
For this purpose, we focus on the scalar-gravity part of supergravity Lagrangian,
which is defined by the frame function $\Omega(z_{i},\bar{z}_{i})$, 
Kahler potential $K(z_{i},\bar{z}_{i})$ and superpotential $W(z_{i})$.
We use $z_i$ to label the chiral superfields including the two Higgs doublets $H_{u,d}$. 

We follow the notation and conventions in \cite{1008.2942},
where the Kahler potential and frame function are given by in unit of Plank mass,
\begin{eqnarray}\label{frame}
K(z_{i},\bar{z}_{i})&=&-3\log(\Omega),\nonumber\\
\Omega(z_{i},\bar{z}_{i})&=&1-\frac{1}{3}\delta_{i\bar{j}}z^{i}z^{\bar{j}}+\cdots.
\end{eqnarray}
Given the explicit form of these functions, 
the scalar potential in the Einstein frame is directly derived from the well-known formula \footnote{Correspondingly, the scalar potential in the Jordan frame is given by $V_{J}=\Omega^{2}V_{E}$.},
\begin{eqnarray}\label{potential}
V_{E}=e^{K}\left(D_{i}W K^{i\bar{j}}D_{\bar{j}}\bar{W}-3W\bar{W}\right)+V_{E}^{(D)},
\end{eqnarray}
where $D_{i}W=\partial_{i}W+K_{i}W$, $V_{E}^{(D)}$ represents the D-term contribution.

The first attempt to embedding Higgs inflation into supergravity was shown in \cite{0912.2718}.
In the light of \cite{0912.2718} there are two important observations.
$1)$, The supergravity version of Eq.(\ref{Lagrangian}) requires a holomorphic function 
$X=-\frac{1}{2}\chi H_{u}H_{d}+\text{h.c}$ in order to reproduce the $\xi$-term.
$\chi$ is a dimensionless coupling constant.
$2)$, The matter content of MSSM is not viable for the Higgs inflation.
By following this line, the authors in \cite{1008.2942} proposed that 
the matter context of NMSSM is a realistic choice,
in which the potential in the NMSSM depends on three complex superfields,
\begin{eqnarray}\label{z}
z_{i}&=&\{S, H^{0}_{u}, H^{0}_{d}\}\nonumber\\
&=&\{se^{i\alpha}/\sqrt{2}, h\cos\beta e^{i\alpha_{1}}/\sqrt{2}, 
h\sin\beta e^{i\alpha_{2}}/\sqrt{2} \}\nonumber\\
\end{eqnarray} 
, as long as the scalar $s$ in the singlet superfiled $S$ can be stabilized at $s=0$ 
and $D$-flat condition $\beta=\pi/4$ is satisfied.
This idea was firstly considered in \cite{0912.2718}, 
but a new $\zeta$-term should be added to the frame function \cite{1008.2942} 
for stabilizing the $s$ scalar.

In summary, the frame function and superpotential for the purpose of Higgs inflation are given by, respectively,
\begin{eqnarray}\label{frame2}
\Omega(z_{i},\bar{z}_{i})&=&1-\frac{1}{3}\left(\mid H^{0}_{u}\mid^{2}+\mid H^{0}_{d}\mid^{2}+\mid S\mid^{2}\right)\nonumber\\
&+&\frac{\zeta}{3} \mid S\mid^{4}-\left(\frac{1}{2}\chi H^{0}_{u}H^{0}_{d}+\text{h.c}\right),\nonumber\\
W(z_{i},\bar{z}_{i})&=&\lambda SH^{0}_{u}H^{0}_{d}+\frac{\rho}{3} S^{3},
\end{eqnarray}
Substituting Eq.(\ref{frame2}) into Eq.(\ref{potential}) gives rise to the potential,
\begin{eqnarray}\label{potential2}
V_{E}=V^{(F)}_{E}=\frac{9\lambda^{2}h^{4}}{(3\chi h^{2}-2h^{2}+6)^{2}},
\end{eqnarray}
for $s=0$ and $\beta=\pi/4$.
Here we have used $W\mid_{s=0}=0$.
For more details about the stabilization of $s$ and angles in Eq.(\ref{z}), 
see \cite{1008.2942, 1004.0712}.
In the region $\chi h^{2}>>1>>h^{2}$, Eq.(\ref{potential2}) 
approaches to $(\lambda/\chi)^{2}$ (in  unit of Plank mass),
which verifies our statements above.
Moreover, in terms of Eq.(\ref{potential2}) one can also verify Eq.(\ref{nr}).

Parameters $\xi$ and $\lambda_{H}$ in Eq.(\ref{Lagrangian}) are related to parameters $\chi$ and $\lambda$ in Eq.(\ref{frame2}) as,
\begin{eqnarray}\label{matching}
\xi=-\frac{1}{6}+\frac{1}{4}\chi,~~~~~~\lambda_{H}(\mu_{I})=\frac{\lambda^{2}}{4}.
\end{eqnarray}
Here $\mu_{I}$ denotes the RG scale corresponding to inflation.
They are constrained by the present cosmological data as follows.
Recall that $V^{1/4}_{E}=\left(24\pi^{2}M^{4}_{P}\epsilon A_{s}\right)^{1/4}$,
where $A_s$ is the amplitude of the power spectrum of the curvature perturbation and $\epsilon=r/16$ in the context of single field inflation.
For $A^{1/2}_{S}\simeq 3.089\times 10^{-5}$ as reported by Plank Collaboration \cite{1303.5076}  one obtains,
\begin{eqnarray}\label{constraint}
\xi \simeq 58789 \sqrt{\lambda_{H}(\mu_{I})}. 
\end{eqnarray}

\section{IV.~Constraints on SUSY mass spectrum}
Until now, we have obtained the value of $\lambda_{H}$ 
below the RG scale $\mu=M_{P}/\xi$ (as shown in Fig.\ref{lambda1} and Fig. \ref{lambda2} ) 
and above the RG scale $M_{P}/\sqrt{\xi}$ (as shown in Eq.(\ref{constraint})) given a choice on $\xi$.
The effective theory at the intermediate scale between  scale $M_{P}/\xi$ and $M_{P}/\sqrt{\xi}$
is described by the SM together with gauginos and squarks. 
Other SUSY particles such as higgsinos and charged Higgs have masses of order $M_{P}/\sqrt{\xi}$ \cite{1008.2942, 1004.0712},
so they should be integrated out at this intermediate energy scale \cite{0812.4946},
especially for the discussion about RGE for $\lambda_{H}$.

The threshold correction $\delta\lambda_{H}$ at scale $\tilde{m}=M_{P}/\xi$ arising from integrating out gauginos and squarks
 can be determined in terms of the  differences between the values of $\lambda_{H}$ below and above scale $\tilde{m}=M_{P}/\xi$,
\begin{eqnarray}\label{threshold1}
\lambda_{H}(\tilde{m}-\epsilon)=\lambda_{H}(\tilde{m}+\epsilon)+\delta\lambda_{H}\left(m_{g_{1}},m_{g_{3}}, m_{\tilde{q}_{i}}\right),\nonumber\\
\end{eqnarray}
where $0<\epsilon<<1$, 
and $m_{g_{1}}$, $m_{g_{3}}$, $m_{\tilde{q}_{i}}$ refers to the gluino mass, bino mass and squark masses, respectively.
As we will see below,  the threshold correction $\delta\lambda_{H}\left(m_{g_{1}},m_{g_{3}}, m_{\tilde{q}_{i}}\right)$ 
can be measured in high precision, 
so it is a new and useful factor to constrain the GUT-scale SUSY mass spectrum.

The value of $\lambda_{H}(M_{P}/\xi+\epsilon)$ in Eq.(\ref{threshold1}) is determined 
by the RGE for $\beta_{\lambda_{H}}$ in the effective theory at the intermediate scale,
with the boundary value at the end of inflation

\begin{eqnarray}{\label{bound}}
\lambda_{H}(M_{P}/\sqrt{\xi})=\lambda_{H}(\mu_{I})+\delta \lambda_{H}\left(m_{\tilde{h}_{u,d}}, m_{s}\right).
\end{eqnarray}
Here, we have ignored the effects due to higgsino-induced operators with mass dimension higher than four, which are at least one order of magnitude smaller than the threshold correction in our case
\footnote{Operators of type $c_{HG}\mid H\mid^{2} F_{\mu\nu}F^{\mu\nu}/M^{2}$  with mass dimension six, as induced by heavy SUSY particles with mass $M$, 
contribute to the leading corrections to the RGEs for SM EW gauge coupling \cite{1402.1476,1308.2627}. As a result, they modify the RGE for $\lambda_H$ indirectly,
the significance of which is determined by the ratio $(m_{h}^{2})_{\text{eff}}/M^{2}$, as seen from the modified beta function to SM gauge coupling, $16\pi^{2}\delta \beta_{g_{i}}\sim c_{HG}g_{i}(m_{h}^{2})_{\text{eff}}/M^{2}$.
For the three stages we set in this paper,
the modification to the RGE for $\lambda_H$ during inflation is the maximal,
with heavy SUSY particle identified as $\tilde{h}_{\mu,d}$.
In contrast, the modification after inflation is only mild.
In our case, $(m_{h}^{2})_{\text{eff}}\leq H$ and $M\sim M_{P}$,
which implies that this modification is smaller than the threshold correction $\delta\lambda_{H}(m_{\tilde{h}_{u,d}}, m_{s})$.}.

The threshold corretion $\delta \lambda_{H}\left(m_{\tilde{h}_{u,d}}, m_{s}\right)$  in Eq.(\ref{bound}) arises from the heavy higgsinos and singlet $s$, which are given by, respectively \cite{1108.6077},
\begin{eqnarray}{\label{values}}
\delta\lambda_{H}\mid_{\tilde{h}_{u,d}}&\simeq&-\frac{1}{6}\cos^{2}(2\beta)\left(\frac{9}{25}g^{4}_{1}+g^{4}_{2}\right)\text{ln}\left(\frac{\mu^{2}}{\tilde{m}^{2}}\right),\nonumber\\
\delta\lambda_{H}\mid_{s}&\simeq& 2\lambda_{H}(\mu_{I})\sin^{2}(2\beta).
\end{eqnarray}
where we have assumed that the A-term $A_{s}$ is smaller than $m_{s}$.
Substituting the stabilized value $\beta=\pi/4$ into Eq.(\ref{values}) 
we find that $\lambda_{H}(M_{P}/\sqrt{\xi})\simeq \frac{5}{2}\lambda_{H}(\mu_{I})$.
As long as the mass scale $m_{\beta}$, which is of order $M_{P}/\sqrt{\xi}$, 
is far larger than the Hubble parameter $H$ 
and the reheating temperature after inflation \cite{0812.4624}, 
$\beta$ rapidly approaches to the stabilized value during and after inflation.

\begin{figure}
\includegraphics[width=0.45\textwidth]{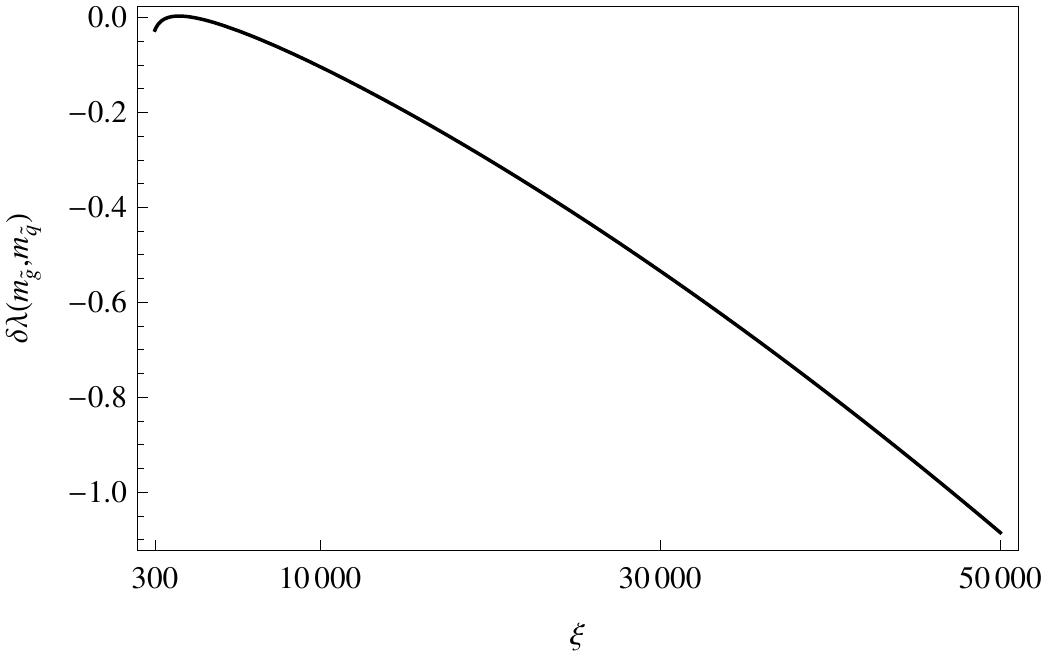}
\caption{Threshold correction $\delta\lambda_{H}\left(m_{g_{1}},m_{g_{3}}, m_{\tilde{q}_{i}}\right)$ in the wino DM model.
We have chosen $m_{t}=172.2$ GeV, $m_{h}=125.5$ GeV and $\alpha_{s}(m_{Z})=0.1184$. 
Similar plot can be obtained for the case with higgsino/singlino DM.}
\label{threshold}
\end{figure}

The matter content of the effective theory at the intermediate scale 
is composed of SM fermions, squarks, gauginos and 
electrically neutral scalars in the Higgs sector.
The beta function coefficients $b_i$ for the SM gauge coupling in this effective theory are given by, respectively,
\begin{eqnarray}{\label{rges}}
b_{1}=\frac{n}{2}, ~~~~
b_{2}=-\frac{3}{2}+\frac{n}{2},~~~~
b_{3}=-\frac{9}{4}+\frac{n}{2},
\end{eqnarray}
where $n=3$ is the number of SM fermion generations.
In terms of Eq.(\ref{rges}) and one-loop RGEs for $\lambda_{H}$ and SM top Yukawa coupling
we show in Fig.\ref{threshold} 
the constraint on the magnitude of $\delta\lambda_{H}\left(m_{g_{1}},m_{g_{3}}, m_{\tilde{q}_{i}}\right)$ 
for the range $300\leq \xi\leq 50000$.
It is shown that the value for $\delta\lambda_{H}\left(m_{g_{1}},m_{g_{3}}, m_{\tilde{q}_{i}}\right)$ as required by Higgs inflation is very sensitive to parameter $\xi$.

In the light of Fig.\ref{threshold}
we conclude that for small non-minimal coupling $\xi \leq 500$,
the threshold correction at scale $M_{P}/\xi$ induced by Higgs inflation is about 
$-0.03\sim-0.02$  and $-0.05\sim -0.04$ in wino and higgsino/singlino DM, respectively.
The uncertainties mainly arise from the uncertainty of top quark pole mass.
In the large $\xi$ region with $\xi\geq 1\times 10^{4}$,
the required threshold correction is $\leq -0.1$, 
which is too large to exclude such models.

\section{V.~Discussion and Summary}
So far we have only discussed the constraint on the threshold correction $\delta\lambda$ 
at scale $M_{P}/\xi$ in the case for wino DM.
In the case for higgsino/singlino DM
the one-loop RGEs for relevant couplings at the intermediate scale are the same as in section IV.
So the conclusion there is also applied to the model of higgsino/singlino DM.
Actually, given the same $\xi$ it is expected that 
the threshold correction for the case with higgsino/singlino DM is larger 
than what is shown in Fig.\ref{threshold}.

One may wonder whether adopting larger $m_{t}$ than $172.2$ GeV can 
reduce the threshold correction in Fig.\ref{threshold} $?$
Fig.\ref{lambda1} and Fig.\ref{lambda2} have shown that $\lambda_{H}(M_{P}-\epsilon)$ becomes more negative 
when one chooses larger $m_{t}$,
and the value of $\lambda(M_{P}/\xi+\epsilon)$, which is positive, will be smaller simultaneously.
Therefore the deviation to the threshold correction is not too obvious to violate our conclusions.

In summary, 
the answer to the question
whether Higgs inflation can be saved by high-scale SUSY critically depends on the magnitude of $\xi$.
For small non-minimal coupling $\xi \leq 500$,
the threshold correction at scale $M_{P}/\xi$ required by Higgs inflation is constrained in high precision,
the magnitude of which is about $-0.03\sim-0.02$  and $-0.05\sim -0.04$ for the wino and higgsino/singlino DM, respectively.
The uncertainties mainly arise from the uncertainty of top quark pole mass.
This amount of threshold correction can be explained,  for example, 
for universal squark masses $m_{\tilde{q}}\simeq \frac{M_{P}
}{6\xi}$ with vanishing $A_t$ terms.
For large $\xi\sim 10^{4}$,
the required threshold correction is smaller than $\simeq -0.1$, 
which excludes such high-scale SUSY.
Finally, we believe our analysis can be applied to Split SUSY, 
although this model is more complicated in compared with what have been discussed here.

\section{acknowledgments}
We would like to thank the referee for valuable suggestions.
The work is supported in part by National Natural Science Foundation of China under grant No. 11247031 and 11405015.

\linespread{1}

\end{document}